\documentstyle[prb,aps,twocolumn,epsfig,floats,amssymb,amsmath,color]{revtex}

\newcommand  {\BondLen}    {b}
\newcommand  {\Rsq}    {\left< R^{2} \right>}

\newcommand  {\cumulant}    {\left< R^{4} \right>/\left< R^{2} \right>^2}

\def\bull{\vrule height .9ex width .8ex depth -.1ex } 

\begin{document}
\draft \title{Equilibration of Long Chain Polymer Melts
in Computer Simulations}

\author{Rolf Auhl$^a$, Ralf Everaers$^a$\thanks{present address:
Max-Planck-Institut f\"ur Physik komplexer Systeme, N\"othnitzer Str. 38, 01187
Dresden, Germany}, Gary S. Grest$^b$,
Kurt Kremer$^a$ and Steven J. Plimpton$^b$}

\address{$^a$Max-Planck-Institut f\"ur Polymerforschung,
  Postfach 3148, D-55021 Mainz, Germany} 
\address{$^b$Sandia National Laboratories, Albuquerque, New Mexico 87185
USA}

\address{
\begin{minipage}{5.55in}
\begin{abstract}\hskip 0.15in
Several methods for preparing well equilibrated melts of long
chains polymers are studied. We show that the standard method in
which one starts with an ensemble of chains with the correct end-to-end
distance arranged randomly in the simulation cell and
introduces the excluded volume rapidly, leads to deformation on
short length scales. This deformation is strongest for long chains
and relaxes only after the chains have moved their own size. Two
methods are shown to overcome this local deformation of the
chains. One method is to first pre-pack the Gaussian chains, which
reduces the density fluctuations in the system, followed by a
gradual introduction of the excluded volume. The second method is
a double-pivot algorithm in which new bonds are formed across a
pair of chains, creating two new chains each substantially
different from the original. We demonstrate the effectiveness of
these methods for a linear bead spring polymer model with both
zero and nonzero bending stiffness, however the methods are
applicable to more complex architectures such as branched and star
polymer.
\end{abstract}
\end{minipage}
\vspace*{-0.5cm}
}

\maketitle

\section{Introduction}\label{sec:intro}
A systematic investigation of the structure-property relations for
polymeric systems by computer simulations requires the preparation
of equilibrated melts of long, entangled chains. For temperatures
well above the glass transition, this can, in principle, be
achieved using sufficiently long molecular dynamics (MD) or Monte
Carlo (MC) simulations. However since the longest relaxation of an
entangled polymer melt of length $N$ scales at least as $N^3$,
giving at least $N^4$ in cpu time, this method is only feasible
for relatively short chain lengths. Depending on the detail of the
model the longest chains which can presently be equilibrated by
brute-force MD or MC simulations are on the order of 2-7
entanglements lengths $N_e$. Using united atom potentials for
polyethylene in which one treats the carbon and its bonded hydrogen
as a single united atom, it is possible to
equilibrate high temperature $(T\agt 450K)$ melts for chains of
the order of approximately $2N_e$ or about 200 
monomers.\cite{paul98,mavrantzas99} For
coarse-grained bead-spring models like the one employed here it is
possible to study longer chain lengths, on the order of up to 500
monomers or approximately $7 N_e$.\cite{puetz00a}
However this reaches the very
limits of present day fastest computers. An increase of cpu speed
by an order of magnitude does even not allow for a doubling of the
chain length. While the present chain lengths are sufficient to
follow the dynamics well into the reptation regime, they are not
long enough to study structure-property relations, such as the
plateau modulus $G_N^o$. This would require chains of order
$10-20N_e$. The situation is even worse for branched polymer melts
or polymers at interfaces, where equilibration times even for
relatively short chains are prohibitively large to use brute force
simulations to produce equilibrated melts.

Fortunately to produce equilibrated samples, there is no need to follow the
slow physical dynamics of entangled polymer melts.  One possibility is to
temporarily turn off the excluded volume interactions and to allow the chains
to pass through each other. In order to obtain a well-defined Monte Carlo
algorithm, it is useful to combine this idea with parallel tempering. While it
has been demonstrated that this is feasible in principle,\cite{bunker01} the
method has so far not proven particularly efficient, the main reason being the
large amount of computer time spent on unphysical Hamiltonians. More
conventional MC algorithms which can be used to equilibrate polymer melts
include reptation moves (generalized slithering snake
algorithms),\cite{vacatello80} configuration bias
algorithms,\cite{depablo92b,siepmann92,dodd94} and concerted rotation
algorithms.\cite{dodd94,dodd93,pant95} Being relatively local, these methods
work best for moderate chain lengths and densities. The complete equilibration
of very long chain melts still requires long runs.  The fastest method, the
generalized slithering snake algorithm, theoretically scales as
$N^2$.  An alternative are algorithms,  which are able to move
large sections of a chain at once.
The prototype for such methods is the pivot
algorithm\cite{lal69,macdonald85,madras88,li95,graessley99b} in which a
monomer is chosen at random and one of the two segments formed by that
partitioning is pivoted rigidly in a random direction about the unit. A
Boltzmann weight is used to determine if the move is accepted or not. This
method is highly efficient for single chains in a continuum, implicit solvent.
Unfortunately, a direct application to dense melts is not feasible since any
large scale conformational change is bound to violate the packing constraints
imposed by the chain excluded volume.  However it is possible to introduce a
double bond crossing algorithm in a way that two new bonds or bridges are
formed across a pair of chains, creating two new chains each substantially
different from the
original.\cite{mavrantzas99,pant95,uhlherr01,Karayiannis02a,Karayiannis02b}
For the bead spring model concerned here this can simply be done by cutting
two bonds and introducing two new bonds as discussed in more detail in
Sec. VII.  For atomistic models, the bond length is much shorter then the
diameter of a monomer, and it is necessary to construct new bridges between
the chain segments.  Karayiannis {\it et
al.}\cite{Karayiannis02a,Karayiannis02b} have shown that by constructing two
trimer bridges between the two properly chosen dimers along the backbone one
can quickly equilibrate long chain polyethylene melts.  For the two chains
involved this is essentially a double-pivot move, where each chain experiences
such a pivot rotation.

In addition to improved equilibration algorithms there is an obvious interest
in methods for generating initial melt configurations which 
are as close to equilibrium as possible.\cite{kremer90,yethiraj90,brown_clarke94,gao95b,mondello98,kroeger99} 
In our previous work on polymer 
melts,\cite{puetz00a,kremer90,kremer95,dunweg98} we
prepared most melts by first generating an ensemble of chains with the correct
end-to-end distance and randomly placing them in the simulation cell.  We
then introduced a weak, non-diverging excluded volume potential in the form of
e.g. a cosine potential, $A(1+\cos(\pi r/2^{1/6}\sigma))$, between non-bonded
monomers, where $r$ is the distance between two monomers and $\sigma$ is the
bead diameter.  The amplitude $A$ was then increased over a short time
interval until the minimum distance between monomers was sufficient to switch
on the Lennard-Jones potential without creating numerical instabilities. In
the first part of this paper we confirm observations by Brown {\it et al.}\cite{brown_clarke94} that this method deforms the (longer) chains. As a
consequence, long chain melts are not as well equilibrated as originally
believed. The deformation turns out to be strongest at short length scales and
completely relaxes away only after a time $\tau_{max}(N)$, which is the time
for a chain to move its own size.\cite{brown_clarke94} This long time is
needed because the local chain-chain topology can only relax by the slow
physical dynamics.  For linear chains, $\tau_{max}(N) \approx
\tau_e(N/N_e)^3$, where the entanglement time $\tau_e=\tau_{\rm Rouse}(N_e)$
is the Rouse time of a chain of length $N_e$ (experimentally the measured
exponent is closer to 3.4 than 3 except for extremely long chains).  This
however was exactly what we tried to avoid by that approach. Not surprisingly,
the longer the chain length $N$, the more severe the effect is. For relatively
short chains, such as those investigated in our previous studies of the
crossover from Rouse to reptation dynamics $(N\le 350)$, the simulations were
run long enough that the results are independent of the starting
state.\cite{puetz00a,kremer90,kremer95} However for longer chains, which
cannot be run long enough compared to the longest relaxation time, this simple
procedure for generating starting states is inadequate.

In the present paper we report results from an extensive effort
to prepare well-equilibrated melts of bead-spring polymers with
chain lengths ranging from $N=350$ to $N=7000$. For this purpose
we use  both approaches outlined above. First we show
how to avoid the local stretching with only minimal computational effort
by a suitable modification of our standard method. Then
we demonstrate the capacity of the double-pivot algorithm to 
dynamically equilibrate melts of medium sized chains of length
up to $N=700$. The paper is organized as
follows: In Sec.~\ref{sec:model} we define the model. 
In Sec.~III, we  present some simple theoretical estimates
of the structure of bead-spring chains with different
intrinsic bending stiffness in dense melts.
In Sec.~IV, 
we  discuss ways to characterize the single chain structure
and extract suitable target functions for long chain melts
from long simulations of short chain melts.
In Sec.~V
we examine carefully the standard procedure used in the past to
prepare  melt configurations. We show that for long chains this
method deforms the chains at short length scales and that these
deformations relax completely only after the chains have moved
their own size. In Sec.~VI, we present the new pre-packing
procedure, which significantly reduces the density fluctuations
particularly for long chains. This combined with a gradual
introduction of the excluded volume, results in well equilibrated
long chain melts in a reasonable amount of cpu time. In Sec.~VI we
describe the double-pivot algorithm and compare results for the
internal dimensions of the chain with those produced from
pre-packing and a gradual introduction of  the excluded volume. In
Sec.~VIII we test the two methods for chains with local bending
rigidity. A brief summary of our main conclusions are given in
Sec.~IX.

\section{Model}\label{sec:model}
We use a coarse grained model in which the polymer is treated as a string
of beads of mass $m$ connected by a spring. The beads interact with a purely
repulsive Lennard-Jones excluded volume interaction,

\begin{equation}\label{eq:ULJ}
  {\rm U_{LJ}}(r) = \left\{
\begin{array}{lcl}
4 \epsilon \left\{ (\sigma/r)^{12} - (\sigma/r)^6 + \frac14 \right\} &
 & r \le r_c \\
0 &  & r\ge r_c
\end{array} \right.
\end{equation}
cutoff at $r_c=2^{1/6}\sigma$. The beads are connected by a finite
extensible non-linear elastic potential (FENE),

\begin{displaymath}
  \rm{U_{FENE}}(r) = \left\{
\begin{array}{lcl}
-0.5kR_o^2 \ln\left(1-(r/R_o)^2\right) & \ r\le R_o \\ \infty & \
r> R_o &
\end{array} \right.
\end{displaymath}
in addition to the Lennard-Jones interaction. The model parameters are the
same as in ref.~\onlinecite{kremer90}, namely $k=30\epsilon/\sigma^2$ and
$R_o=1.5\sigma$ unless otherwise noted. The temperature $T=\epsilon/k_B$. The
basic unit of time is $\tau=\sigma(m/\epsilon)^{1/2}$. We performed constant
volume simulations of monodisperse melts at a segment density
$\rho=0.85\sigma^{-3}$. The temperature was kept constant by coupling the
motion of each bead weakly to a heat bath with a local friction
$\Gamma$. Unless otherwise specified $\Gamma=0.5m/\tau$. The equations of
motion were integrated using a velocity Verlet algorithm with a time step
$\Delta t=0.012\tau$. The average bond length is $<\BondLen^2>^{1/2}
=0.97\sigma$. The polymer melts studied consisted of $M$ chains of length $N$
beads. The chain lengths studied varied from $N=25$ to $7000$. The number of
chains are each system is specified in Table~\ref{tab:systems}.

Equilibration algorithms for polymer melts which follow the physical dynamics
encounter a strong growth of the necessary equilibration times with chain
length.  Per unit of time $\tau$, the simulation of a melt of $M$ chains of
length $N$ requires a computational effort of $10^2 M\times N$ particle
updates. Since the longest relaxation time is of the order $\tau_{max}\approx
\tau_e (N/N_e)^3$,  equilibration of our largest system with $M=80$
and $N=7000$ would require 
$10^2 M\times N (\tau_e/\tau) (N/N_e)^3\approx 8\times
10^{17}$ particle updates.  Assuming a typical performance of $2.5\times10^5$
particle updates per second, which is typical for this model on a single
processor, the required cpu time for the equilibration is about 
$10^5$ years.  On a modern parallel cluster such as the DEC alpha CPlant
cluster at Sandia, the times are somewhat less.  On 64 processors, we get
about 20 steps per second for our largest system, which means that the total
time for the chains to move on average their own size is approximately 1600
years. Increasing the numbers of processors helps somewhat but since the
efficiency of the computation decreases as the number of processors is
increased for fixed system size, this type of brute force approach for long
chain melts is clearly not feasible.

In  our previous studies, we considered only fully flexible
chains. However to demonstrate the effectiveness of the
methodology, we also include a nonzero bending stiffness, which is
modelled by

\begin{equation}\label{eq:Ubend}
  U_{bend}(\theta) = k_\theta \left(1-\cos \theta \strut\right)
\end{equation}
where 
$\cos \theta_i =(\hat {\bf r}_i-\hat {\bf r}_{i-1})\cdot(\hat{\bf r} _{i+1}-\hat {\bf r} _i)$, where
$(\hat {\bf r}_i-\hat {\bf r}_{i-1})$ is the
unit vector in the bond direction.

\section{Chain characteristic in the melt}  
\label{sec:StructureEstimates}
\begin{figure}[t]
  \begin{center}
  \includegraphics[angle=0,width=0.95\linewidth]{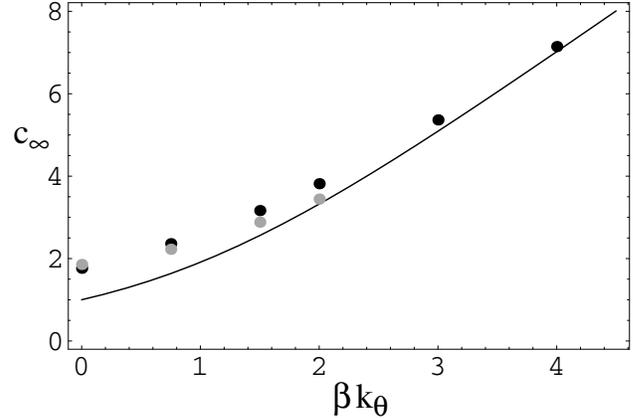}
  \caption{Estimates of the effective stiffness $c_\infty$ of our chains as a
  function of the strength $k_\theta$ of the bending potential
  Eq.~(\protect\ref{eq:Ubend}).  Eq.~(\protect\ref{eq:cos_theta_of_k_theta})
  (solid line) only accounts for the bending energy, while
  Eq.~(\protect\ref{eq:cos_theta_NNNEV}) ($\bullet$) also considers excluded
  volume interactions between next-nearest neighbor monomers along the
  chain. The gray dots indicate estimates based on the analysis of simulation
  data for equilibrated melts of chain length $N=200-350$ for $k_\theta=0$ and
  $50-100$ for $k_\theta>0$.   }
  \label{fig:EffectiveStiffness} \end{center}
\end{figure}

In dense polymer melts excluded volume interactions between different parts of
a given polymer chain are screened.  Since in our model there are no explicit
intrachain interactions beyond those between neighboring bonds, the single
chain structure should be well characterized by the expectation value $\langle
\cos \theta \rangle$ of the bond angle $\theta$. For example, the knowledge
of $\langle \cos \theta \rangle$ is sufficient to calculate an estimate for
the mean-square end-to-end extension $\langle R^2\rangle$ of chain segments of
length $N$ monomers and $n=N-1$ bonds,

\begin{equation}\label{eq:r2_of_N}
\langle R^2\rangle(n) = n \BondLen^2 \left( \frac{1+\langle \cos \theta \rangle}{1-\langle \cos \theta \rangle} -
\frac1n
\frac{2 \langle \cos \theta \rangle \left(1-\langle \cos \theta  \rangle^n\strut\right)}
{\left(1-\langle \cos \theta \rangle\strut\right)^2}
\right)
\end{equation}\label{eq:cinftyG}
where the asymptotic prefactor is referred to as
\begin{equation}
c_\infty =\frac{1+\langle \cos \theta \rangle}{1-\langle \cos \theta \rangle}.
\end{equation}

The simplest estimate of $\langle \cos (\theta) \rangle$ only takes
the bare bending energy Eq.~(\ref{eq:Ubend}) into account. Neglecting
small variations in the bond length around the mean value
$\BondLen=0.97\sigma$, one finds:

\begin{eqnarray}\label{eq:cos_theta_of_k_theta}
\langle \cos \theta \rangle & = &
  \frac{\int_0^\pi \sin \theta \cos \theta \exp\left(-\beta U_{bend}(\theta)\right)d\theta}
       {\int_0^\pi \sin \theta              \exp\left(-\beta U_{bend}(\theta)\right)d\theta}\nonumber\\
&=&
\frac{1 + e^{2\beta k_\theta}\left(\beta k_\theta-1 \right)  + \beta k_\theta}
  {\left(e^{2 \beta k_\theta} -1 \right) \beta  k_\theta},
\end{eqnarray}
where $\beta=1/k_BT$.
Eq.~(\ref{eq:cos_theta_of_k_theta}) is represented as a solid line in 
Fig.~\ref{fig:EffectiveStiffness}. 
However this result underestimates the effective chain stiffness,
since the chain cannot fold back. This effect can be accounted for
approximately by modelling the chains as non-reversal-random-walks
(NRRWs) with excluded volume interactions between next-nearest
neighbors along the backbone. As a function of the bond angle
$\theta$ their distance is given by $l_\theta=|\vec r_{i+1}-\vec
r_{i-1}| = \BondLen[2(1+\cos \theta_i)]^{1/2}$ so that

\begin{eqnarray}\label{eq:cos_theta_NNNEV}
\langle \cos \theta \rangle & =  &
  \frac{\int_0^\pi \sin \theta \cos \theta
           \exp\left(-\beta (U_{bend}(\theta)+U_{LJ}(l_\theta))\right)d\theta}
       {\int_0^\pi \sin \theta
           \exp\left(-\beta (U_{bend}(\theta)+U_{LJ}(l_\theta))\right)d\theta}
\end{eqnarray}
which has to be evaluated numerically 
(see the black dots in Fig.~\ref{fig:EffectiveStiffness}).
For $k_\theta=0$ this gives $c_\infty=1.76$ compared to
$c_\infty=1$ obtained from Eq.(~\ref{eq:cinftyG}) and in good
agreement with simulation results for chains of length $20\le
N\le 350$.  This value  
is slightly smaller than the value $c_\infty=2.06$ obtained at the
$\Theta$-point for the same Lennard-Jones interaction truncated at
$2.5\sigma$ in an implicit, continuum solvent.\cite{graessley99b}

Ensembles of single chains which obey Eq.~(\ref{eq:r2_of_N}) can be
generated by simple sampling. This is particularly simple, if the
allowed bond vectors are distributed evenly over a solid angle with
$\theta\le \theta_{max}$.  In this case

\begin{equation}\label{eq:cos_theta_of_theta_max}
\frac{c_\infty-1}{c_\infty+1}=
\langle \cos \theta \rangle_{\theta_{max}}
=\left( \cos \theta_{max}/2\right)^2
\end{equation}
so that the appropriate value for $\theta_{max}$ can be easily determined
for any desired effective stiffness.

\section{Target functions and suitable ways of characterizing
 single chain structure}\label{sec:target}

The estimates of the chain structure presented in the previous
section were obtained within an effective single chain picture. In
reality, the formal characterization of a polymer melt is a
complicated many-body problem. For example, we have completely
ignored the influence of packing effects on the chain
conformations. As was recently shown\cite{abrams02} the local melt
structure very sensitively depends on the ratio of bond length to
effective excluded volume of the beads. This effect is not well
represented by the value of the overall chain extension. Here we
use our standard bond length/diameter ratio and focus on the
single chain properties, however the methods discussed below 
apply to systems with different parameters as well. In the present
case, sufficiently long 
MD or MC simulations represent the only ``ab initio'' method which
allows to take into account all interactions.

\begin{figure}[t]
  \begin{center}
    \includegraphics[angle=0,width=0.95\linewidth]{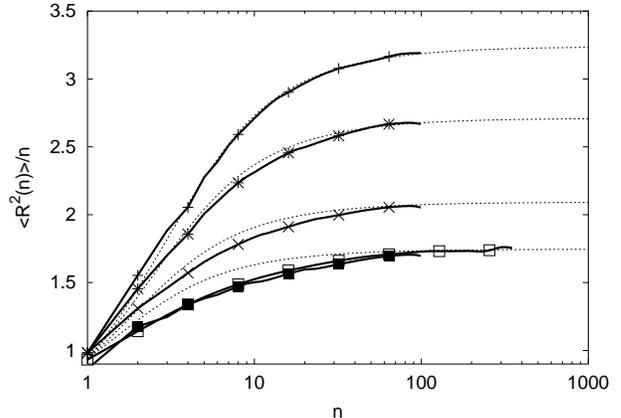}
  \end{center} 
\caption{Mean square internal distances from long MD runs for
  chain length $N=100$ and $350$ respectively for four values of
  the intrinsic bending stiffness: $k_\theta=0$ ($\square$), $k_\theta=0.75$
  ($\times$), $k_\theta=1.5$ ($\ast$), and $k_\theta=2.0$ ($+$). We refer to
  these data sets as ``target functions''. For comparison, they are included
  as thick black lines in subsequent, corresponding plots.  The dashed lines
  show to a fit of Eq.~(\ref{eq:r2_of_N}) to the
  asymptotic behavior. The corresponding values for $c_\infty$ are included in
  Fig.~\protect\ref{fig:EffectiveStiffness}.}  
\label{fig:target}
\end{figure}

In order to characterize the chain conformations in the melt we mainly rely on
mean-square internal distances $\Rsq(|i-j|,N)$ averaged over all segments of
size $n=|i-j|$ along the chains, where $i<j\in [1,N]$ are monomer indices.
In addition we characterize deviations from Gaussian
statistics on short length scales using the ratio $\cumulant$.
Figures~\ref{fig:EffectiveStiffness} and
\ref{fig:target} show a comparison of simulation results for equilibrated
melts and our naive estimates. 
The results from the melts have been run sufficiently long that the
chains have moved several times their own size. For $N=350$, $k_\theta=0$,
the run time  $T_t=2.6\times 10^6\tau$.\cite{puetz00a}
Clearly, the description of chains in a melt as
simple freely jointed chains is too naive. Nevertheless, the deviations are
not large and the measured and estimated values for $c_\infty$ agree quite
well. In particular, the simulation data show neither unexpected features nor
significant finite chain length effects.

The rest of the present paper deals with algorithms which try
to circumvent the slow physical equilibration path. These algorithms
will be judged according to two criteria: their capacity to reproduce
the target functions and their performance. We mostly concentrate
on the fully flexible case. In the end, we come back to the stiffer
systems as a kind of ``blind test''.

\section{Standard procedure for Preparing Polymer Melts}
\label{sec:old}
Our standard method for preparing melt conformations is
as follows:

\begin{figure}[tb]
  \begin{center}
    \includegraphics[angle=0,width=0.95\linewidth]{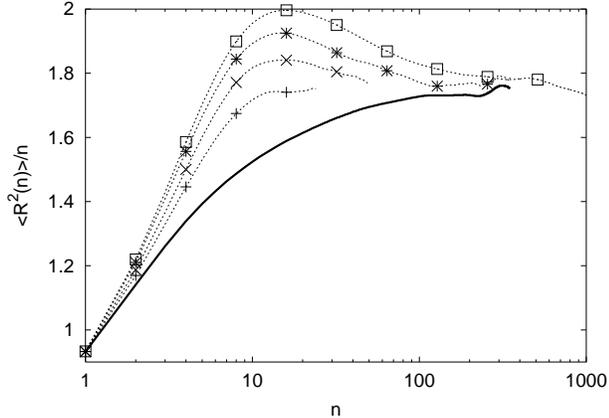}
  \end{center}
  \caption{Mean square internal distances after a fast push-off for
randomly packed phantom chains of length $N=25\ (+)$, $50\ (\times)$,
$350 \ (*)$ and $7000 (\square)$. The thick line is our target function from
Fig.~\protect\ref{fig:target}.}
  \label{fig:fastpushoff}
\end{figure}

\begin{figure}[tb]
  \begin{center}
   \includegraphics[angle=0,width=0.95\linewidth]{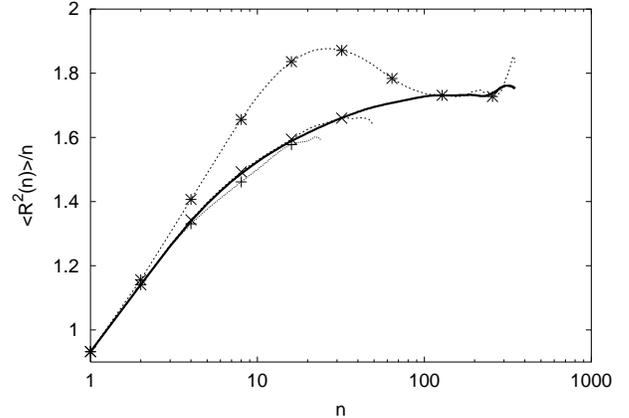}
  \end{center} \caption{Mean square internal distances for the systems shown
  in Fig.~\protect\ref{fig:fastpushoff} after an additional MD relation of
  $10^4\tau$ which corresponds to $\tau_{\rm Rouse}(N=90)$. Symbols as in
  Fig. ~\protect\ref{fig:fastpushoff}. The thick line is our target function
  from Fig.~\protect\ref{fig:target}.}  \label{fig:fastpushoff_plusMD}
\end{figure}

\begin{enumerate}
\item Start from an ensemble of chains with the correct end-to-end
distance $R^2(N)=\BondLen^2 c_\infty(N-1)$
on large length scales. For a known $c_\infty$
this step is trivial. 
\item Arrange the chains randomly in the simulation box. 
\item Introduce a weak, non-diverging excluded
volume potential. A convenient form for the soft potential is a
cosine potential,
\begin{displaymath}\label{eq:cos}
  {\rm U_{soft}}(r) = \left\{
\begin{array}{ll}
A(1+\cos (\pi r/r_c))  & r\le r_c \\
0 & r\ge r_c
\end{array} \right.
\end{displaymath}
between non-connected monomers with $r_c=2^{1/6}\sigma$. The
initial amplitude of $A=4\epsilon$ was linearly increased to a
final values of $A\ge 100\epsilon$ (``push-off'') over a short
time interval of 10-20 $\tau$. In the final state the inter
monomer distances are large enough to allow one to switch to the
LJ potential without creating numerical instabilities.  During
this phase, we often use a stronger coupling to the thermostat by
increasing the friction constant to $\Gamma=2.0m/\tau$. 
In addition, the
velocities of all monomers are set to zero every $50$ time steps.
This facilitates monomers that are strongly overlapping to
separate. We refer to this method in which the excluded volume
interactions are introduced fairly rapidly as ``fast push-off.''
\item Relax the system  with the full LJ-potential by  a short
MD run.
\end{enumerate}

Figure~\ref{fig:fastpushoff} shows that this procedure actually
deforms the chains so that the melt is {\em not} equilibrated
after step 3. The deformations are strongest on short lengths. With
increasing $N$ only the amplitude but not the position of the
deformation along the chain changes. The monomer displacements
during step 3 are too small to affect the conformations on large
scales, allowing us to tune them to preselected values.  Since the
largest length scales also take the longest time to relax, one
might hope that the MD relaxation in step 4 can be significantly
shorter than for a starting conformation of, say, completely
stretched chains. To be more specific, a naive expectation is that
chain segments of length $n\le N$ equilibrate on time scales
comparable to the Rouse/reptation time of chains of length $n$
{\em independent} of $N$. Figure~\ref{fig:fastpushoff_plusMD}
compares the result of equilibration runs of length
$T_t=10^4\tau\approx\tau_{max}(N=90)$ to the target function in
Fig.~2.  As expected, the internal
distances measured for chains of length $N=25$ and $N=50$
coincide. However segments with $n\alt 50$ of longer chains for
$N=350$ are still far from being equilibrated after $10^4\tau$.
Even after a time $t\approx \tau_{\rm Rouse}(N)\sim
\tau_e(N/N_e)^2\sim 10^5\tau$  (not shown) the chains of length
$N=350$ were not fully equilibrated. Rather the local equilibration is
completed only after the chains have moved their own size,
$t>\tau_{max}(N)$.  In lattice simulations it is sometimes possible to avoid
the slow reptation dynamics by allowing the chains to cut through each
other.\cite{shaffer95,shaffer96} In this case $\tau_{max}$ is given by a
Rouse-time and scales as $N^2$. In off-lattice bead-spring models topology
conservation is the result of energetic barriers which prevent chain
crossing. 
Since these
barriers are a result of the microscopic interaction potentials,
they are difficult to manipulate without affecting the local chain
structure. Hence Rouse-like relaxation mechanisms due to 
barrier crossing dominate
reptation only for extremely long (and therefore inaccessible) chain
lengths. Attempts to circumvent these barriers using parallel tempering
\cite{bunker01} have met with limited success at least as
far as efficiency gains are concerned.

\section{Optimized Procedure for Polymer Melt Conformations}\label{sec:rolf}
The methods presented in this section aim at a more careful
implementation of the idea underlying our standard procedure: the
local build-up of the characteristic melt packing for chains with
the correct large length scale statistics. In this section we show
that we can achieve this goal by first pre-packing the phantom
chains and a subsequent {\em slow} and improved push-off scheme
for the excluded volume. Both steps are needed to produce well
equilibrated melts in a reasonable amount of cpu time. Applying
only one does not achieve this objective. All results in this section
are averaged over five independent realizations of the packing and
push-off procedures. The system sizes studied are listed in 
Table~\ref{tab:systems}.

\begin{table}
\begin{tabular}{r|r}
M&N\\
\hline
500&100\\
250&200\\
120,200&350\\
500&500\\
200 &700\\
200&1400\\
80& 3500\\
80& 7000
\end{tabular}
\caption{\label{tab:systems}
System size studied: Number of chains $M$ and chain length $N$.}
\end{table}

\subsection{Prepacking of phantom chains}
\label{sec:prepacking} 
\begin{figure}[t]
  \begin{center}
    \includegraphics[width=0.95\linewidth,angle=0]
                    {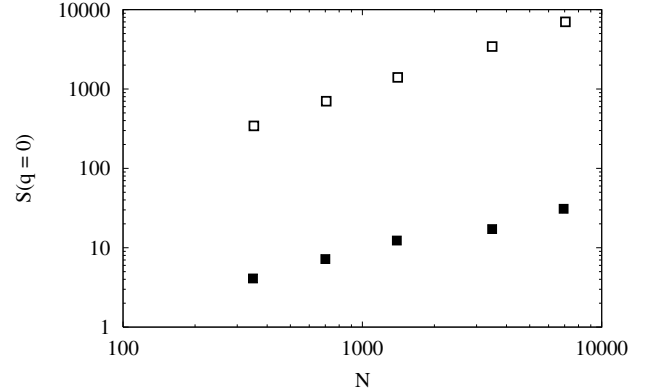}
    \caption{Density fluctuation of randomly distributed $(\square)$ and
      pre-packed Gaussian chains $(\bull)$ for
     various chain lengths. By pre-packing the density
      fluctuations are  about 1\%  for $N=350$ and 0.5\%
      for $N=7000$ of that for randomly distributed chains.}
    \label{plot_dens_fluct_rd_and_pp}
  \end{center}
\end{figure}

The chain deformations created by the standard
procedure are due in part to the large local density fluctuations
in a ``melt'' of randomly overlapping NRRW chains. As the
chain length increases, these fluctuations increase in size as
shown in Fig.~\ref{plot_dens_fluct_rd_and_pp}. In this section we
describe a dynamic Monte-Carlo-like pre-packing procedure for
phantom chains, which  significantly reduces these density
fluctuations.

As a merit function we have chosen the fluctuations $E=\langle
n^2\rangle-\langle n\rangle^2$ in the number $n_i$ of particles found in a
sphere of radius $d$ around particle $i$. In a homogeneously dense system
$E=0$. We perform a zero temperature Monte Carlo (MC) optimization 
in which all
moves which decrease the density fluctuations are accepted and all moves which
increase the density fluctuations are rejected.  In the course of the packing
run $d$ is decreased from $d=4\sigma$ to $2\sigma$. We use a linked cell
structure in order to calculate the $n_i$ efficiently. Choosing $d>\sigma$
effectively increases the range of the excluded volume correlations.

All MC moves change the positions or orientations of entire chains
which are treated as rigid. This pre-packing procedure therefore
does not affect the single chain statistics, which by construction
already have the correct end-to-end distance $R$ as well as the
targeted internal distance distributions. We consider five types
of MC moves:

\begin{description}
\item[Translation] of individual chains in a random directions.
\item[Rotation] of individual chains by random angles around
random axes through their centers of mass. 
\item[Reflection] of individual
chains at random mirror planes going through the center of mass. 
\item[Inversion] of individual chains at their centers of mass. 
\item[Exchange] of two chains preserving the center of mass positions.
\end{description}
Typical run times were of the order of days on individual workstations
and therefore negligible.

\begin{figure}[t]
  \begin{center}
    \includegraphics[angle=0,width=0.95\linewidth]{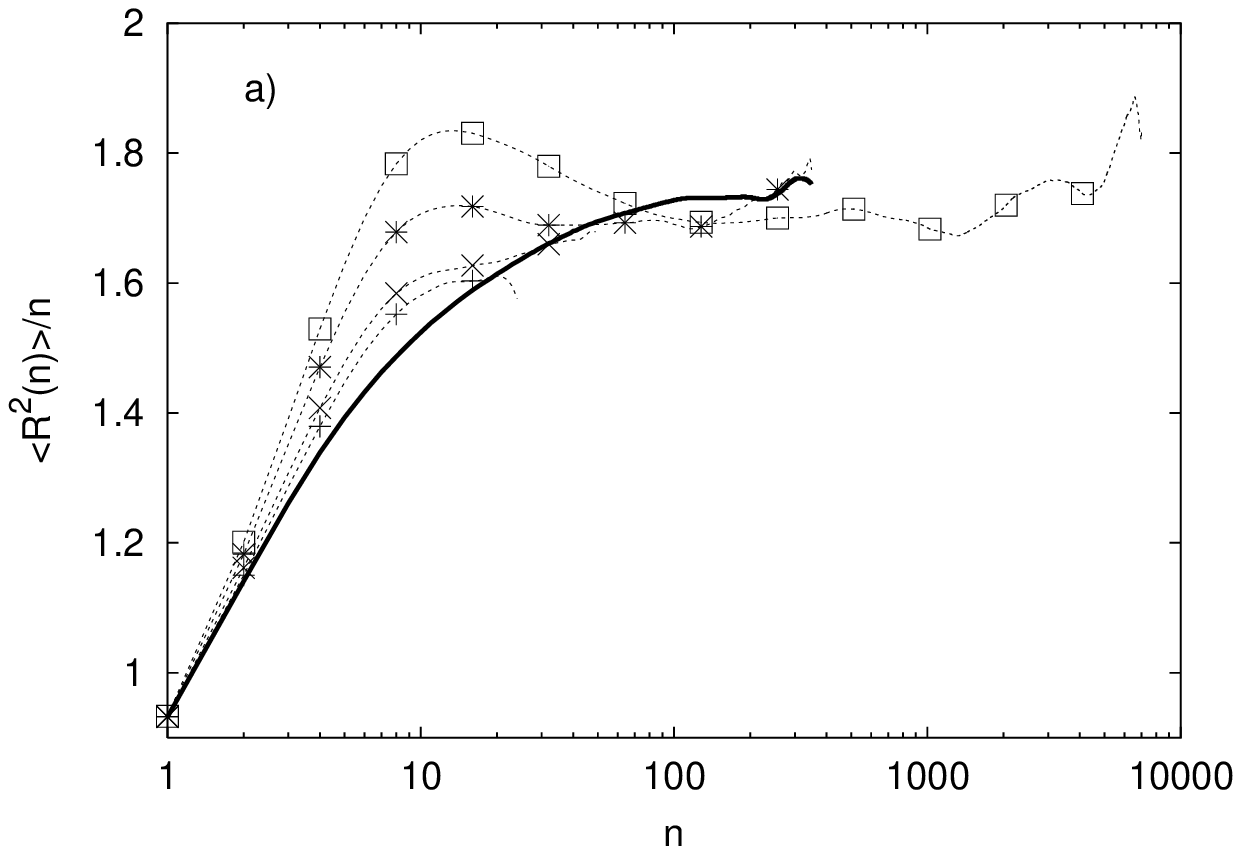}
   \includegraphics[angle=0,width=0.95\linewidth]{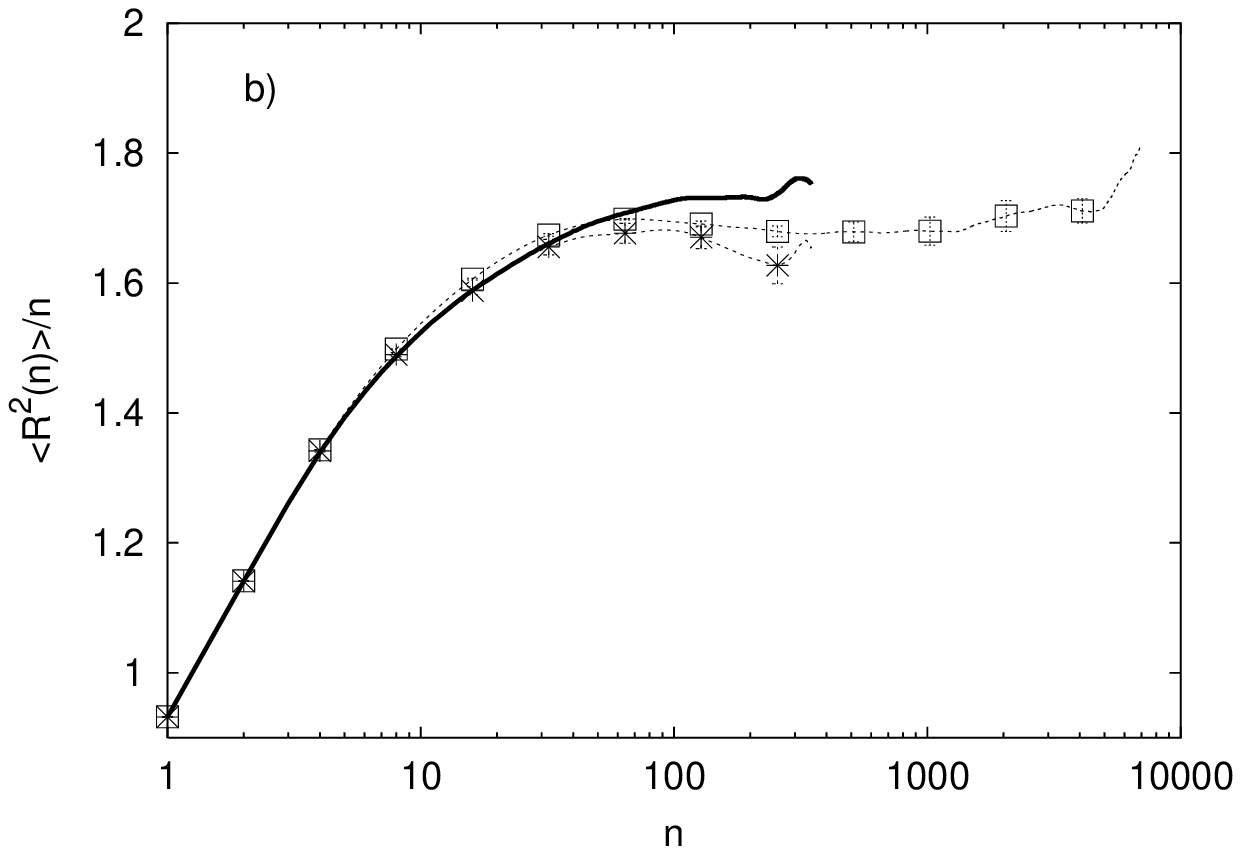}
  \end{center} \caption{Mean square internal distances for pre-packed systems
  after a (a) fast push-off and (b) slow push-off for chains with
  $k_\theta=0$. Symbols as in
  Fig. ~\protect\ref{fig:fastpushoff}. The thick line is our target function
  from Fig.~\protect\ref{fig:target}.}  \label{fig:prepack_plus_pushoff}
\end{figure}

The amplitude of the density fluctuations is drastically reduced as the
packing proceeds. The final result is best assessed by the $q\rightarrow 0$ 
limit of
the system structure function $S(q)$.  Figure~\ref{plot_dens_fluct_rd_and_pp}
shows that the pre-packing reduces the density fluctuations to one percent of
the initial value. Applying the fast push-off procedure outlined in Sec.~V
to these pre-packed states, we see by comparing
Figs.~\protect\ref{fig:fastpushoff} and
\ref{fig:prepack_plus_pushoff}a that the local deformations
are eliminated for short chains $(N\alt 50)$ but not for long
chains. For long chains, there is a significant reduction in the
local stretching but not enough to avoid the necessity of
subsequent long  MD runs, which again would need to be of the
order of the disentanglement time $\tau_{max}(N)$.

\subsection{Slow push-off}
\label{sec:slowpushoff}
In order to further reduce the chain deformations we have modified
the push-off procedure in three ways:

\begin{enumerate}
\item Replace the cos-potential Eq.~(\ref{eq:cos}) with a
  force-capped-Lennard-Jones-potential,~\cite{brown_clarke94}. 
\item Keep the full excluded volume between next-nearest neighbor
monomers to preserve to non-reversal-random-walk structure (see
sec.~\ref{sec:StructureEstimates}).
\item Increase the push-off time.
\end{enumerate}

Force-capping~\cite{brown_clarke94} can either be defined directly via a
maximum force $F_{\mathrm max}$ or implicitly via a distance $r_{fc}$ where
${\rm U_{LJ}}^\prime(r_{fc})\equiv F_{\mathrm max}$.  At larger separations,
the interaction is defined by Eq.~(\ref{eq:ULJ}), at smaller distances the
force becomes constant so that the interaction potential grows only linearly with
$(r - r_{fc})$:

\begin{displaymath}
{\rm U_{FCLJ}}(r) = \left\{
\begin{array}{cc}
(r - r_{fc})*{\rm U_{LJ}}^\prime(r_{fc}) + 
{\rm U_{LJ}}(r_{fc}) &   r <r_{fc} \\
{\rm U_{LJ}}(r) &    r \ge r_{fc}
\end{array} \right.
\end{displaymath}
In the present case, $r_{fc}$ is gradually reduced from
the Lennard-Jones cut-off radius $r_c=2^{1/6}\sigma$ to 
$0.8\sigma$ which is significantly smaller than the 
relevant interparticle distances. 

Force-capping has the advantage that the soft potential
systematically approaches the true potential.
In contrast, the cos-potential
overcompensates the missing singularity at the origin by a fast
rise of the repulsive interactions close to the cut-off distance.
This leads to slight differences in the monomer packing. As a
result, the chain conformations are slightly perturbed after
switching to the full LJ-potential.

We have increased the push-off time in order to introduce the excluded volume
interactions quasi-statically.  In practice, one has in this context to worry
about two issues: (i) How slow is slow enough?  (ii) What is the equilibrium
statistics for polymers with force-capped interactions?  Concerning the first
point, we have chosen a push-off time of $5000\tau$, which is of the order  
$\tau_{\rm Rouse}(50)$.  This covers the typical 
subchain regime, up to which the intrachain distances still were disturbed 
by a fast push-off.
The second question is more difficult to answer. If we simply switch off the
repulsive monomer-monomer interaction, we find $\Rsq=0.94\sigma^2N$, while the
effective stiffness $c_\infty\approx1.7$ of our chains in the melt is due to
local bead packing effects. Thus a  slow push off starting from a potential
for which the end-to-end distance is  an ideal
random walk is also dangerous. 
For bead-spring chains these effects can easily be accounted for
by treating the chains as
NRRWs by always keeping the full LJ excluded
volume between next-nearest neighbors along the chain.  
For atomistic polymer models, it is necessary to extend this to
larger distances along the chain (``pentane effect'').\cite{brown_clarke94}
Once this adjustement is made the push
off procedure cannot be ``too slow''and the chain size varies
only very weakly with the maximum excluded volume force.

\begin{figure}[t]
  \begin{center}
  \includegraphics[angle=0,width=0.95\linewidth]{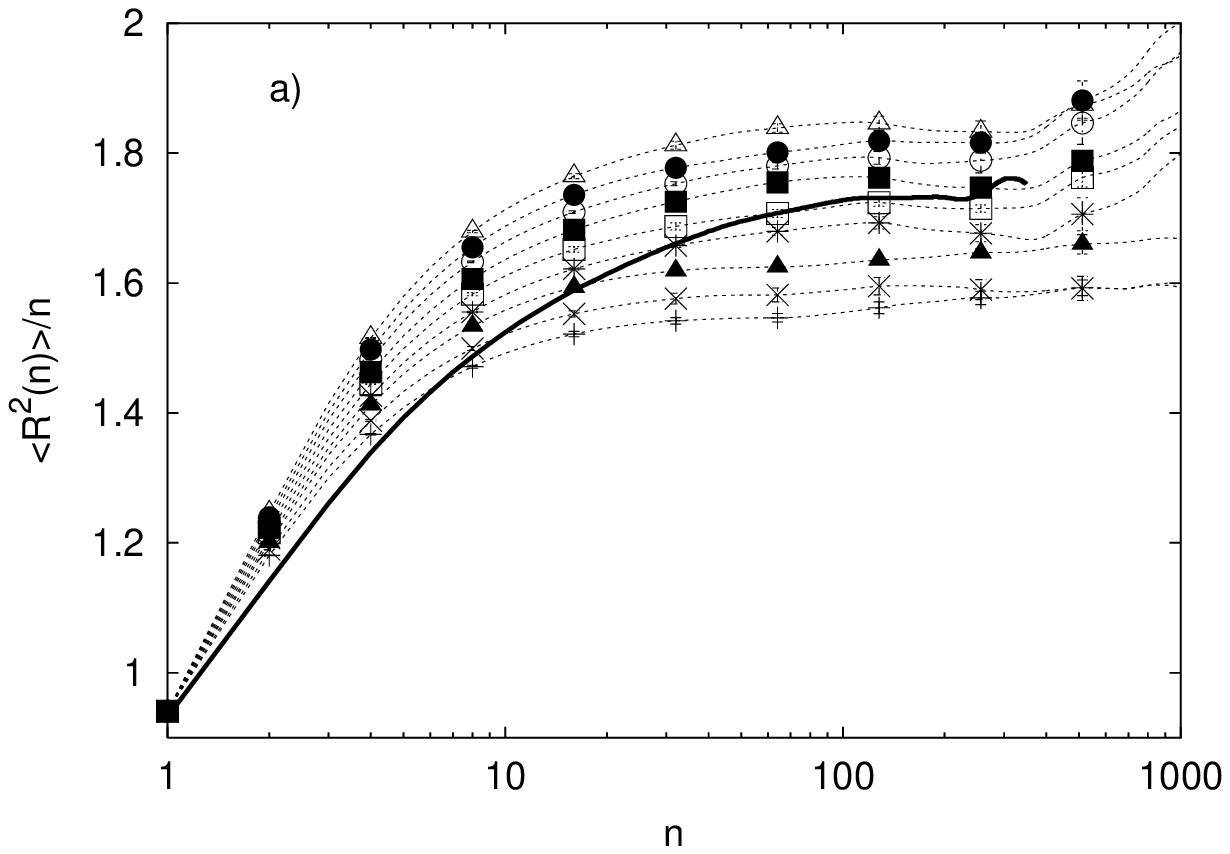}
  \includegraphics[angle=0,width=0.95\linewidth]{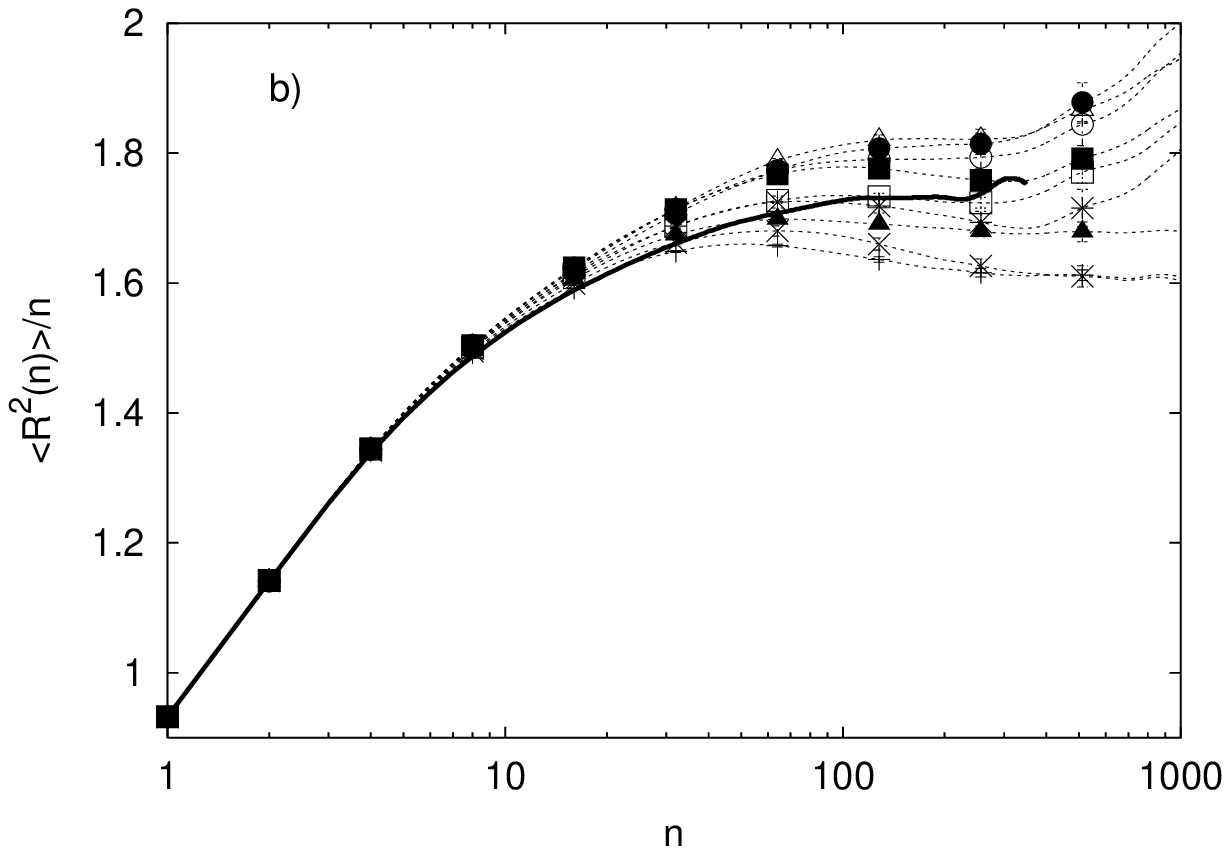}
  \end{center} \caption{Influence of the effective stiffness used to generate
  the initial sets of random walks on the final configurations for systems
  with $k_\theta=0$. The figure shows mean square internal distances (a)
  before and (b) after pre-packing and slow push-off.  From top to bottom the
  curves represent results for chains with length $N=3500$ which were set up with a stiffness
  varying by $-4\%,\ -2\%,\ 0\%,\ +2\%,\ +4\%,\ +6\%,\ +8\%,\ +10\%$  and
  $+12\%$ relative to
  our initial guess of $c_\infty=1.7$. The thick line is our target function
  from Fig.~\protect\ref{fig:target}.}  \label{fig:excl}
\end{figure}

Figure~\ref{fig:prepack_plus_pushoff}b shows that the new procedure works
quite well. Independent of the total chain length $N$ the chains are no longer
stretched on intermediate scales and the large scales remain unaffected by the
introduction of excluded volume interactions. As an additional check, we have
varied the stiffness of the initially generated random walks over a range of
about 20\% (Fig.~\ref{fig:excl}a).  Fig.~\ref{fig:excl}b compares the internal
distances after pre-packing and slow push-off and confirms our previous
observations.  The chains are fully equilibrated on short scales, where the
results are independent of the initial conditions. In contrast, internal
distances beyond $n=100$ remain practically unaffected.  We therefore conclude
that our new method allows the preparation of well-equilibrated melts of
extremely long chains within reasonable computational effort, provided the
correct asymptotic chain stiffness $c_\infty$ 
is known from a careful extrapolation of
simulation results for well-equilibrated short and medium chain length melts.
For our largest system of $M=80$ chains of length $N=7000$, an equilibration
time of $5000\tau$ takes about one month on a single processor or about 8
hours on the 100 processor DEC alpha cluster. Thus even with moderate
computational means it is possible to equilibrate very large systems using a
combined pre-packing and slow push-off procedure.


\section{Double-Pivot-MD Hybrid Algorithm}\label{sec:pivot}

The present section deals with the problem of accelarating
the dynamic equilibration of a long chain polymer melt.
We describe a double-pivot-MD hybrid algorithm which performs
this task significantly faster than simple MD relaxation.
In contrast to the methods discussed in the previous section,
the dynamic equilibration of a dense polymer melt 
does {\em not} require independent knowledge of the effective $c_\infty$.
However, the preparation of locally equilibrated samples
can significantly reduce the required relaxation times.


In the original pivot algorithm\cite{lal69,macdonald85,madras88,li95} one
choses a monomer at random and one of the two segments formed by that
partitioning is pivoted rigidly in a random direction about the selected
monomer. The pivot algorithm is extremely efficient when applied to isolated
chains in an implicit solvent.  However, in a melt or even in a solution at
moderate density, pivot moves applied to a {\em single} chain are bound to be
rejected, since they lead to strong excluded volume interactions with other
chains.

An attractive alternative are Monte Carlo moves involving at least two chains,
which change the connectivity within the melt in a way that the overall
packing of beads remains (almost) unaffected. Such algorithms were first
introduced for lattice models\cite{mansfield82,olaj82} and recently extended
to off-lattice 
models.\cite{mavrantzas99,pant95,uhlherr01,Karayiannis02a,Karayiannis02b}
The efficient equilibration usually comes at the expense of a certain amount
of polydispersity which is introduced into the samples, though 
recently Karayiannis\cite{Karayiannis02a,Karayiannis02b} have
overcome this limitation using  double-bridging 
Monte Carlo moves.  
For the bead-spring polymers studied here, the algorithm is straightforward to
implement since the bond length $l$ is comparable to the excluded volume
parameter $\sigma$. This certainly is a unique situation, as for many bead
spring coarse grained models of specific chemical species, the above ratio is
not as close to one. 
For example in the case of a
united atom model for polyethylene where the bond length $l=1.54\AA$ is
significantly smaller then $\sigma\approx 
4.0\AA$, it is necessary
to construct two trimer bridges between the two properly chosen dimers
along the backbone.\cite{Karayiannis02a,Karayiannis02b} 
For the bead spring model studied here, complex bridging moves are
not necessary. For this reason we refer to the method as the double-pivot
algorithm since it is much closer to the pivot algorithm used for
single chains in an implict, continuum solvent.

In order to maintain a monodisperse melt we search for a pair of
spatial neighbor monomers $i$ and $j$ on different chains or on
the same chain, which
happen to also be the same distance from one end of their
respective chain. Then one can replace two bonds along the
original chains with two new bonds or bridges across the pair of
chains. The total change in energy $\Delta E$ is local and
consists of the sum of the energies of the two new bonds minus the
energies of the two previous bonds as well as the difference in the
bending energy,
which involves the sum of  the energies of four
new angles minus the energies of the four previous angles
for the case $k_\theta\ne 0$.
The move is accepted
by a standard Metropolis criterion, namely with a Boltzmann weight
$\exp{(-\Delta E/k_BT)}$ if $\Delta E>0$ and $1$ if $\Delta E \le
0$. As in the original pivot algorithm, the double-pivot algorithm
generates new chains which are substantially different from the
original two chains since as many as $N/2$ monomers of a chain are
replaced by monomers from the other chain. This results in
immediate large changes in the end-to-end distance and radius of
gyration.

We have implemented the double-pivot algorithm into both our shared memory and
distributed memory MD codes. The details of the implementation and a
discussion of its computational efficiency will be published
elsewhere. Briefly in our shared memory code, every $3-5$ time steps, we
randomly chose a monomer $i$ and check its non-bonded nearest neighbors to
determine if any satisfy the condition that they are equidistant from the end
of their chain.  If so, the energy change $\Delta E$ of the double-pivot move
is determined and the move accepted or rejected on the basis of the Metropolis
criterion. If the move is accepted, monomers and/or the connectivity table are
re-labeled depending on which of the two codes is used and the MD simulation
is continued.  If the move is rejected or no suitable pair is identified, a
new monomer is chosen at random and the process repeated. If no suitable pair
is generated after searching a specified fraction of monomers (usually $2-5\%$
depending on the chain length $N$), we continue with the MD simulations for
another $3-5$ steps and repeat the procedure.  On the distributed memory code,
the procedure is similar except that each processor randomly choses a monomer
$i$ and checks its non-bonded nearest neighbors on the same processor to see
if they satisfy the condition that they are equidistant from the end of their
chain.  If no suitable non-bonded neighbor is found, another monomer $i$ is
randomly selected. The process is continued until a specified fraction of the
momomers, typically 50\%, are tested.  To facilitate determining the distance
from the free end, each chain carries an extra label starting from $1$ at
either end to $N/2$ at the center.  Because each processor searches a unique
region of space independently, it is possible to have multiple successful
moves each time the procedure is applied. The search is restricted to sets of
monomers on the same processor to avoid the need to communicate changes in
chain topologies between processors. This restriction also prevents two or
more processors from performing simultaneous swaps that could energetically
conflict with each other. While this restriction means (slightly) less swaps
are considered, this is more than outweighed by the parallelism, e.g.  up to
$P$ swaps take place in one time step, where $P$ is the number of processors.
With the standard parameters of the FENE potential, $k=30\epsilon/\sigma^2$,
the acceptance rate is quite low. A way to improve the acceptance rate is to
start with a somewhat lower value, $k=10-15\epsilon/\sigma^2$ and gradually
increase $k$ during the course of the simulation.  This change has little
effect on the overall structure of the chains but increases significantly the
acceptance rate. Using our distributed memory code on 27 DEC alpha
processors for a system of $M=500$ chains
of length $N=500$, the number of successful exchanges for $10^5\Delta t$ with
$k_\theta=0$ was approximately $1750$ for $k=30\epsilon/\sigma^2$ compared to
$8500$ for $k=20\epsilon/\sigma^2$ and $49500$ for
$k=10\epsilon/\sigma^2$. 

\begin{figure}[tbh]
  \begin{center}
  \includegraphics[angle=0,width=0.95\linewidth]{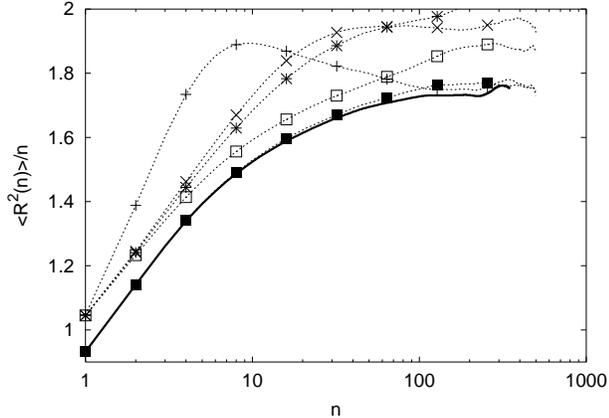}
  \end{center} 
\caption{Equilibration of mean square internal distances using
  combined double-pivot-MD relaxation
  following a fast push-off for 500 chains
  of length $N=500$.  Results for $t=0 \ (+)$ just after the push-off,
  $6\times 10^3\tau \ (\times)$, $1.2\times 10^3\tau \ (*)$, 
  $6\times 10^4\tau \ (\square)$ and
  $1.2\times 10^5\tau \ (\bull)$. In order to increase 
  the acceptance rate of the
  pivot moves we used a reduced spring constant 
  $k=10 \epsilon/\sigma^2$ for $0 < t < 6\times 10^4\tau$ and 
  $k=20 \epsilon/\sigma^2$ for $6\times 10^4\tau<t<9.6\times 10^4\tau$ 
  before switching to
  $k=30 \epsilon/\sigma^2$ for $9.6\times 10^4\tau<t<1.2\times 10^5\tau$.
 } \label{fig:MDPivot}
\end{figure}

To demonstrate the
effectiveness of the double-pivot algorithm, we have applied it to 
a melt of $M=500$ chains of length $N=500$
generated using the standard fast push-off procedure outlined in Sec.~V.
Figure \ref{fig:MDPivot} shows the internal distances as
a function of time during a double-pivot/MD relaxation.
Note that the characteristic ``bump'' in the internal distance
function does not simply decay.  Rather, Figure~\ref{fig:MDPivot} shows that
the (seemingly equilibrated) largest scales are first swollen, before the chain
dimensions decay homogeneously on all scales. Equilibration for the reduced
spring constant $k=10\epsilon/\sigma^2$ was achieved after about
$6\times10^4\tau$ or 10 accepted pivot moves per monomer.  
The penalty for reducing the spring constant to $k=10$ is an increase
of the mean-square bond length by $7\%$, while
 the chain stiffness $c_\infty$ remains unaffected.
In subsequent runs, the spring constant $k$ is increased to
slowly ``reel in'' in the chains. This required an
additional $6\times10^4\tau$.  We made no attempt to optimize the number of
steps for each value of $k$. Nevertheless, the required equilibration times
are considerably shorter than the time $\tau_{max}(500)\approx 4\times 10^6\tau$
for the chains to move their own size in a pure MD simulation.

These observations illustrate the close analogy between the pivot algorithm
for single chains and the double-pivot algorithm for dense polymer
melts. As pointed out by Madras and Sokal\cite{madras88}, it is useful to
distinguish between {\em decorrelation} and {\em equilibration} in discussing
the performance of such algorithms.  Pivoting is extremely efficient in
decorrelating large length scales, {\em provided} the initial conformation is
properly equilibrated. Up to small corrections, the number of pivot moves
needed to decorrelate the end-to-end distance of isolated chains in an
implicit solvent is ${\cal O}(10)$ {\em independent} of chain length. Since
the same holds for subchains of arbitrary length, correspondingly more moves
are required to decorrelate higher order (Rouse) modes.  In other words, the
decorrelation proceeds from large to small scales.  However, the same is {\em
not} correct for the equilibration of a perturbed initial configuration. In
this case it is necessary to run the system up to the decorrelation time of
the {\em shortest} perturbed length scale in order to equilibrate the chains
on {\em all} length scales.  For single chain studies aiming at global
properties Madras and Sokal~\cite{madras88} therefore recommend to apply the
pivot algorithm to equilibrated initial configurations generated by other
techniques. In the present case, the combination of pre-packing with a slow
push-off at least partially fulfills these requirements, since the chains are
locally equilibrated.

Since equilibration requires that each monomer comes 
close enough to suitable exchange partners, one may ask whether
of ${\cal O}(10)$ exchanges per monomer independent
of $N$ is sufficient or whether
the exchanges are simply being made back and forth between a 
limited number of states.  To estimate whether enough independent
configurations are being sampled, consider the time for a monomer to
diffusive a typical distance between monomers with the same index
$(\rho/(N/2))^{-1/3}$ between successful pivot moves. 
If we assume that monomers diffuse with 
the typical $\langle \delta r^2 \rangle = \sqrt{k_BT b^2 t/\zeta}$ 
Rouse dynamics, this corresponds to 

\begin{equation} 
t_{piv}=\left(\frac{ (N/2)}{\rho}\right)^{4/3} 
\frac\zeta{k_BT b^2} 
\end{equation} 
which is equivalent to the Rouse time of a chain of length 
$N_{eff}=(N/2)^{2/3}$.
For our longest chains $N=7000$, $N_{eff}\approx 236$. The Rouse
time for a chain of length $N=236$ is approximately $7\times10^4\tau$,
which is still much less than the time needed to make ${\cal O} (10)$
exchanges per monomer even if we used the reduced spring
constant $k=10$. Thus only for much longer chains or more
complex architectures does one have to worry about sufficient
independent configurations being sampled by this hybrid
double-pivot-MD algorithm.

\section{Local bending rigidity}

\begin{figure}[tb]
  \begin{center}
    \includegraphics[angle=0,width=0.95\linewidth]{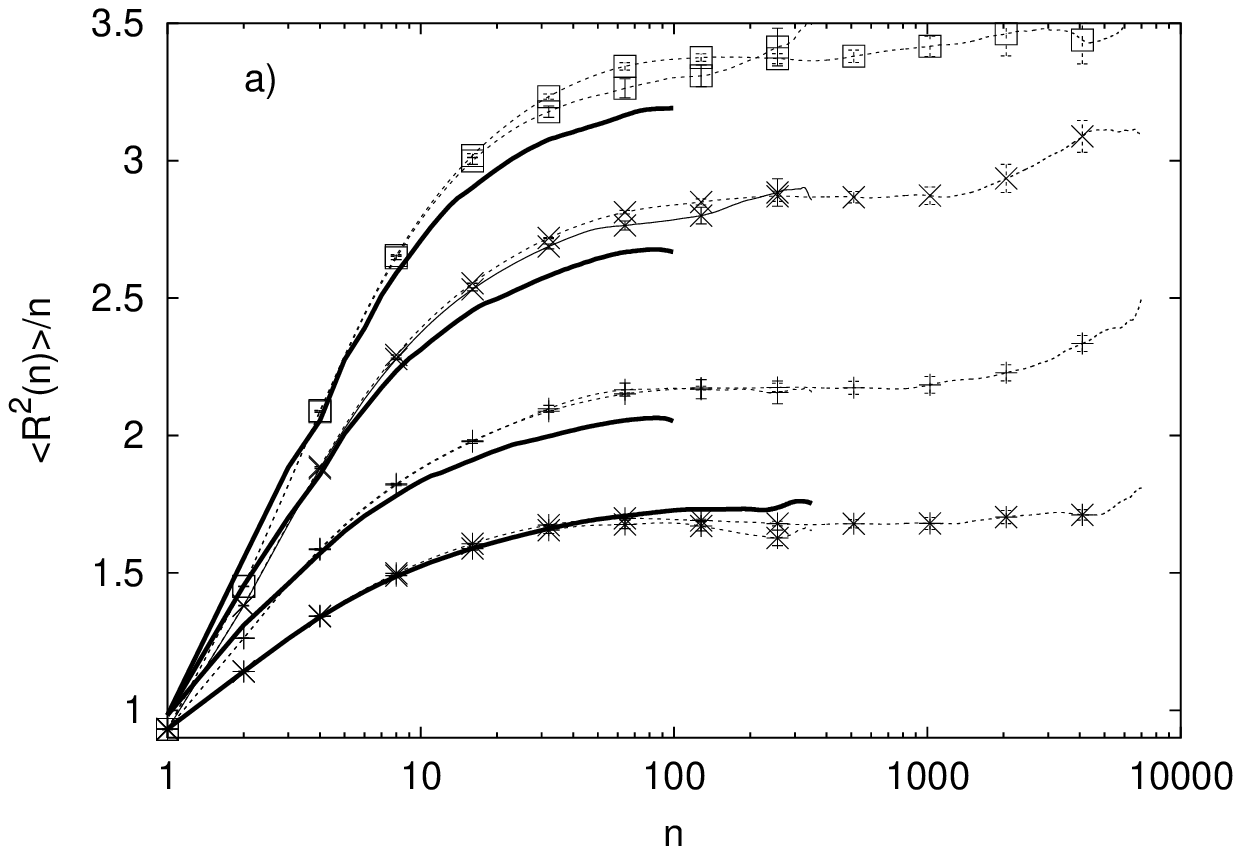}
    \includegraphics[angle=0,width=0.95\linewidth]{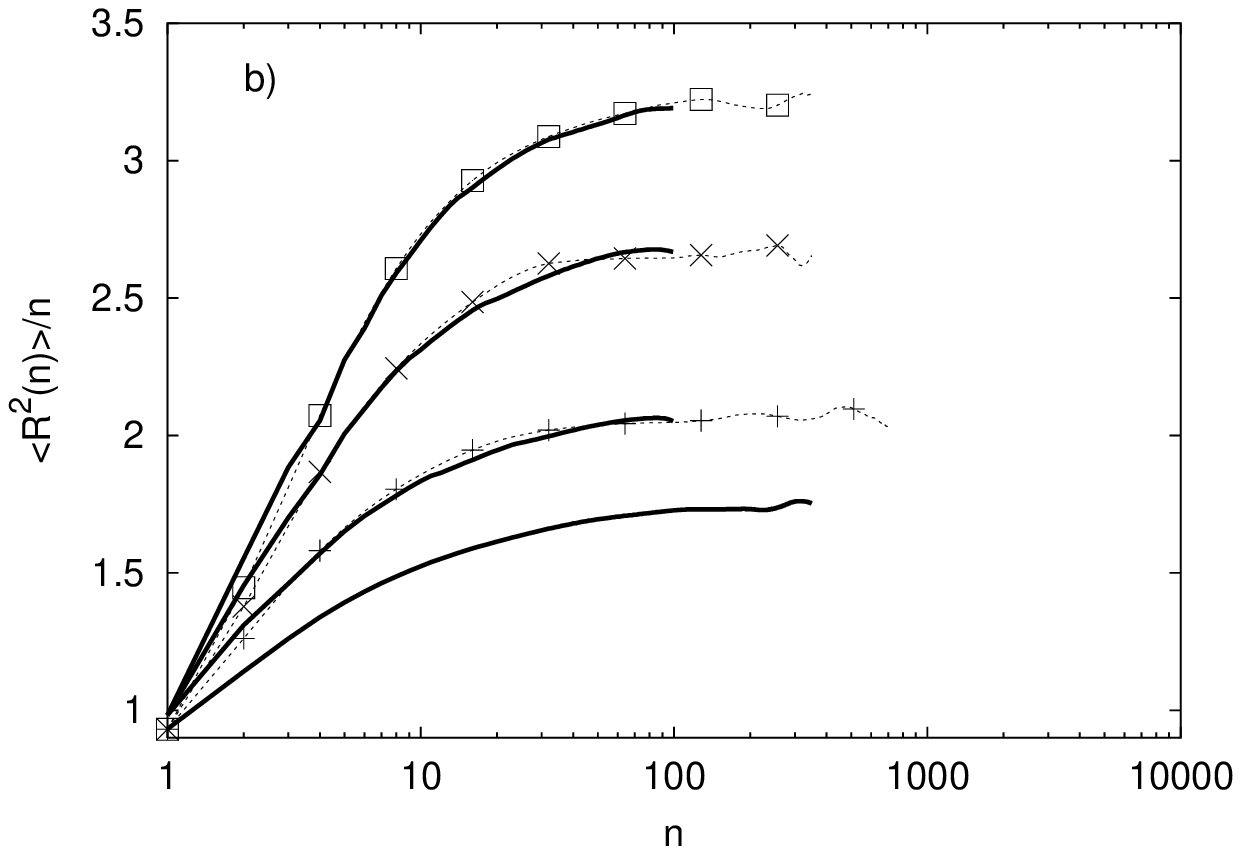}
  \end{center}
  \caption{Mean square internal distances for chains with local bending 
     rigidity (a) after pre-packing and  slow push-off and (b) after an
     additional double-pivot-MD relaxation of length $1.08\times 10^5\tau$.
     Results for $k_\theta =0\ (*)$, $0.75\epsilon\ (+)$, 
     $1.5\epsilon\ (\times)$, and $2.0\epsilon\ (\square)$. We did not
apply the additional double-pivot-MD relaxation to our fully flexible
systems, which are well-equilibrated after pre-packing and slow
push-off.
     The thick lines are our target functions from 
     Fig.~\protect\ref{fig:target} which were generated {\em a posteriori}
     for chains with intrinsic stiffness.
}
  \label{fig:stiffness}
\end{figure}

As a last point we present results for chains with intrinsic stiffness
Eq.~(\ref{eq:Ubend}). In section~\ref{sec:target} we referred to this case as
a ``blind test,'' because the starting states were generated based on our
simple estimate Eq.~(\ref{eq:cos_theta_NNNEV}) for the effective chain
stiffness in the melt and {\em before} the brute-force MD runs for the target
functions shown in Fig.~\ref{fig:target} were completed. As can be seen in
Fig.~\ref{fig:EffectiveStiffness}, our original estimates were slightly too
large. 
A comparison with the target functions after the pre-packing and slow push-off
phase (Fig.~\ref{fig:stiffness}a) shows that as for fully flexible chains
the correct chain statistics is reproduced on short length scales. However, we
are now in a situation comparable to the
situtaion presented in  Fig.~\ref{fig:excl} where the large
length scales are not fully equilibrated, because the initial chains were
generated with a slightly incorrect effective stiffness.

\begin{table}
\begin{tabular}{l|r|r|r}
		& $k=10$	& $k=20$	& $k=30$\\
\hline
$k_\theta=0.0$  & 24800		& 3900		& 840\\
$k_\theta=0.75$ & 27000		& 4200		& 894\\
$k_\theta=1.5$  & 21200		& 3500		& 770\\
$k_\theta=2.0$  & 15500		& 2700		& 600\\
\end{tabular}
\caption{Influence of the strength of the FENE spring constant
$k$ on the number of successful double-pivot exchanges per 
$10^5$ time steps for a system of 200 chains of length
$N=350$ on 27 DEC alpha processors.}
\label{tab:PivotAcceptance}
\end{table}

Subsequently, we ran the combined double-pivot-MD simulation for $1.08\times
10^5\tau$ with $k=10$ for the first 4 millions steps, $k=20$ for the second
$4$ million steps and $k=30$ for the last million steps for a system of $200$
chains of $N=350$ for $k_\theta=1.5$ and $2.0$ and $N=700$ for
$k_\theta=0.75$.  The resulting mean squared end-to-end distance were in
excellent agreement with the target functions generated by brute force MD
simulations for shorter chains.  Table~\ref{tab:PivotAcceptance} shows an
example of the number of successful exchanges per $10^5$ steps for the system
of $200$ chains of length $N=350$ monomers for three values of the spring
constant $k$ for four values of $k_\theta$ for our distributed memory code run
on 27 DEC alpha processors.  As seen from this table, the procedure used gives
approximately $9-12$ successful exchanges for $k=10$, $2$ successful exhanges
for $k=10$ and $0.1$ for $k=30$.  Fortunately there is little difference in
the local structure of the chain for $k=20$ compared to $k=30$, though as seen
from Fig.~8, the chain is expanded for $k=10$. Thus there is a definite
tradeoff between reducing the bond spring constant $k$, which increases the
number of exchanges while at the same time increasing the mean squared bond
length.  Note that for longer chains lengths, the length of the run and/or the
number of processors would have to be increased so that the number of
successful exchanges is on the order of ${\cal O}(10)$. This limits the
double-pivot/MD hydrid method to moderate chain lengths.

\section{Conclusions}

In this paper, we have discussed the preparation of equilibrated melts
of  long
chain polymers in computer simulations. The interest in reliable
algorithms for this task is due to two problems: (i) the
prohibitively long relaxation times encountered in brute force MD
simulations and (ii) the difficulty to predict
the chain statistics on large scales (i.e. the effective chain stiffness
$c_\infty$) on the basis of microscopic intra- and interchain
interactions.

Our results can be summarized as follows:

\begin{itemize}
\item It is insufficient to judge the quality of the equilibration
from the statistics of the chain end-to-end distances or radius of
gyration.
The first just measures one length, while the latter gives
a non specific average over all internal distances. As a
much more sensitive criterion we measure mean square internal
distances on {\em all} length scales from the monomer size up to the
contour length of the chains. 

\item Our standard method to rapidly introduce excluded volume
interactions between randomly assembled Gaussian chains with the
correct overall statistics (``fast push-off") {\em fails}, when
judged according to this improved criterion. The fast push-off
introduces significant chain distortions on length scales of the
order of the tube diameter. They 
can only vanish when the local melt topology is equilibrated.
In MD simulations this is only possible via reptation dynamics
so that the proper equilibration requires runs whose
length exceeds the disentanglement time of
the chains. 
To overcome this problem we have introduced two different methods.

\item We have modified our standard approach by reducing the
density fluctuations in the assembly of Gaussian chains
(``pre-packing") and by introducing the excluded volume
interactions in a quasi-static manner (``slow push-off"). As
before, this method requires prior knowledge of $c_\infty$. Tests
for bead-spring models with chain lengths up to $N=7000$
$(N/N_e={\cal O}(100))$ show that suitable push-off times are the
Rouse times of the characteristic chain lengths of the overshoot.
In the present case this time is of the order of a few
entanglement times, {\em independent} of $N$.

\item We have applied a ``double-pivot"
algorithm which is inspired by the highly efficient pivot algorithm for {\em
single} chains and the double-bridging method for dense systems. For
intermediate chain lengths, this is a viable method 
to equilibrate melts in a reasonable amount of cpu time, particularly     
when one has {\em no} prior
knowledge of $c_\infty$.  

\end{itemize}
The combination of MD relaxation, double-pivot and slow push-off
allows the efficient and controlled preparation of equilibrated
melts of short, medium and long chains respectively.  
While our results were obtained for an off-lattice bead spring model with
chain lengths up to $N=7000$ beads, the methods should work as
efficiently for lattice polymer models such as the bond
fluctuation model.\cite{carmesin88,deutsch91} 
The pre-packing and gradual introduction of the
excluded volume are also applicable to united and explicit atom
simulations.
Furthermore, the methods are applicable to polydisperse melts and
to branched polymers, including long chain branching, comb and
star polymers. The double-pivot and double bridging algorithms 
has also been used to
equilibrate long end-tethered, polymer brushes in contact with a
melt of long chains.\cite{Karayiannis02b,grest02} 
For these more complex chain architectures,
in which the chains are not necessarily Gaussian and there is no
{\it a priori} way to know the conformation of the chains,
algorithms like the double-pivot and double-bridging algorithms are
presently
the only possible way  to equilibrate  systems faster than
extremely long MD or MC runs.

\bibliographystyle{prsty}
\bibliography{macros,ref13}


\end{document}